\documentclass[a4paper,twoside]{article}

\usepackage{epsfig}
\usepackage{subcaption}
\usepackage{calc}
\usepackage{amssymb}
\usepackage{amstext}
\usepackage{amsmath}
\usepackage{amsthm}
\usepackage{multicol}
\usepackage{multirow}
\usepackage{pslatex}
\usepackage{makecell}
\usepackage{apalike}
\usepackage{xurl}
\usepackage{tabularx}
\usepackage[bottom]{footmisc}
\usepackage{SCITEPRESS}     % Please add other packages that you may need BEFORE the SCITEPRESS.sty package.

\usepackage{natbib} % citing authors by names with \citet{}, can be used as \usepackage[numbers]{natbib} for numbered citations

\begin{document}

\title{Rethinking Certification for Higher Trust and Ethical Safeguarding of~Autonomous Systems}

% \author{\authorname{Dasa Kusnirakova\orcidAuthor{0000-0002-5341-902X} and Barbora Buhnova\orcidAuthor{0000-0003-4205-101X}}
\author{\authorname{Dasa Kusnirakova and Barbora Buhnova}
\affiliation{Faculty of Informatics, Masaryk University, Brno, Czech Republic}
% \affiliation{\sup{2}Department of Computing, Main University, MySecondTown, MyCountry}
\email{\{kusnirakova, buhnova\}@mail.muni.cz}
}


\abstract{With the increasing complexity of software permeating critical domains such as autonomous driving, new challenges are emerging in the ways the engineering of these systems needs to be rethought. Autonomous driving is expected to continue gradually overtaking all critical driving functions, which is adding to the complexity of the certification of autonomous driving systems. 
As a response, certification authorities have already started introducing strategies for the certification of autonomous vehicles and their software.
But even with these new approaches, the certification procedures are not fully catching up with the dynamism and unpredictability of future autonomous systems, and thus may not necessarily guarantee compliance with all requirements imposed on these systems. 
In this paper, we identified a number of issues with the proposed certification strategies, which may impact the systems substantially. For instance, we emphasize the lack of adequate reflection on software changes occurring in constantly changing systems, or low support for systems' cooperation needed for the management of coordinated moves. Other shortcomings concern the narrow focus of the awarded certification by neglecting aspects such as the ethical behaviour of autonomous software systems. 
The contribution of this paper is threefold. 
First, we discuss the motivation for the need to modify the current certification processes for autonomous driving systems. 
Second, we analyze current international standards used in the certification processes towards requirements derived from the requirements laid on dynamic software ecosystems and autonomous systems themselves. 
Third, we outline a concept for incorporating the missing parts into the certification procedure.}

\onecolumn \maketitle \normalsize \setcounter{footnote}{0} \vfill

\section{\uppercase{Introduction}}

According to the most recent estimates, the majority of human driving will be replaced by autonomous vehicle (AV) technology by the year 2050 \citep{litman2022}. Besides the expected benefits of increased road safety, significant cost savings and reduced energy consumption and pollution \citep{dia2020}, the introduction of self-driving vehicles into public places creates new challenges for safeguarding the mobility ecosystem as a whole in order to ensure vehicles' safe operation. Therefore, alongside the quickly emerging autonomous software advancements, it is critical to develop regulatory frameworks that would be designed to adapt to the technological changes, so that the safety risks are minimized.

Regulations inevitably have to evolve to keep up with technological progress, more so in the face of the increasing software intensity of autonomous-driving systems. This has already occurred in the history of vehicle certification methods. Initially, when vehicles were made only from mechanical components (brakes, tyres), the driving function and all decision-making was in hands of the driver. In this case, classic certification approaches were sufficient. However, following the introduction of ABS\footnote{Anti-lock Braking System; a safety system used on land vehicles activated in case of a skid in order to allow the driver to maintain more control over the vehicle.} and other systems with a greater level of complexity, it became clear that the classical approach was insufficient in assessing all safety-relevant aspects because of the extensive number of potential testing scenarios. As a consequence, process- and functional-oriented safety audits were introduced, one of which is the Annex 6 of the UN Regulation no.79 \citep{un-reg-79}.

Future road vehicles are expected to gradually overtake more critical driving functions, until manual driving might be fully replaced. This will shift the focus and the responsibility from the driver towards the system installed in the AV. With that, the importance and complexity of the electronic control systems utilized in vehicles will continue to increase. This will also significantly increase the number of possible scenario variations. However, the testing phase performed in the same way as for conventional vehicles, that is, verifying the system based on a predefined set of tests, will be able to thoroughly examine only a limited subset of all safety areas and scenarios \citep{kalra2016}. Besides that, autonomous systems, i.e. \textit{"systems changing their behaviour in response to unanticipated events during operation"} \citep{watson2005autonomous}, among which AVs clearly belong, rely on automatic software updates at runtime in order to adjust to specific environment or context changes \citep{deco2021}, or to improve AI components of an AV in general \citep{cert-eu-report2020}. Again, the traditional certification procedures are not designed to promptly deal with this situation. They do not assume any, or at most a limited number of changes in already certified systems. But just like software engineering does not end with the deployment of the system, future AVs will require novel technological as well as legal approaches for their thorough quality control even during runtime in order to ensure public road safety. 

The challenges of the AV certification go even beyond the safety of autonomous ecosystems. Studies have shown that trust plays an essential role in the adoption of automated systems \citep{trust-cioroaica2019}, from both the societal (people are willing to accept and use the systems) as well as technological (interactions between communicating systems during runtime are trustworthy) point of view \citep{survey-trust-mng2022}. However, continual technological progress of AV exceeding the boundaries of previously defined safety requirements hinders trust formation in these systems \citep{trust-cioroaica2020}, and certification, generally perceived as a trust-building mechanism, is failing to provide sufficient legal guarantees in its current state. It is, therefore, necessary to adapt the certification methods to the envisioned technological advancements and facilitate the adoption process of AV into society.

In this paper, we first explore the necessity to improve the present certification methods for autonomous driving systems, and second, we evaluate recently published standards against the identified requirements for future autonomous systems. Taking into consideration the expertise from social computing as well as the characteristics of dynamic autonomous ecosystems in which such systems operate, we outline a concept for incorporating the missing parts into the certification procedure. To this end, we formulate the requirements for the future certification procedures of autonomous systems, select relevant standards and evaluate them against the defined requirements.
% Furthermore, we analyze the suitability of current certification approaches to these requirements and we list examples in which the existing certification falls short.

% Therefore, we believe currently established certification procedures for conventional vehicles may no longer be appropriate for the future usage. 
The rest of the paper is organized as follows. Section 2 presents related work in the field of certification of autonomous driving systems. Section 3 lists the characteristics and specifies the certification context for autonomous systems, based on which we define the requirements for standards used for the certification of future autonomous systems. In Section 4, we describe the research methodology and select certification standards for evaluation. Section 5 presents the evaluation results. In Section 6, we outline the suggestions for improvements, needed for the standards to reflect the specifics of future autonomous (eco-)systems.

% regular software updates will not concern only newly manufactured cars for sale, but will be needed also for vehicles that have already been produced.

% Such increased dynamics in the software updating process will also increase the probability for a potentially faulty SW being installed in the vehicle (malfunctioning either intentionally - a virus, hijack, or unintentionally - bug not revealed by tests). 

% * Current certification process is sufficient for static environments.
% * But physical and one-time testing, typically at the time of production, is no longer sufficient to assess an AV's performance in a variety of real-world scenarios, and dynamically changing contexts. 
% * Moreover, despite technical, procedural and environmental aspects, morality and ethical behaviour of the smart agents need to be taken into consideration and somehow reflected in the certificate, too.
% Therefore, there is an emerging need for modification of the current certification concept in order to reflect the actual state in a rapidly changing environment and to serve as a trustworthy source of information for all further decision-making processes within such a dynamic autonomous ecosystem.

\section{\uppercase{Related Work}}
\label{sec:relwork}

The question of the suitability of the current vehicle certification processes for the future has already been to some extent discussed in the literature. However, to the best of our knowledge, no paper has yet highlighted the importance of balancing both trust and ethics in the process of certifying autonomous systems.

% critics on current certification
\citet{bonnin2018} has pointed out that certification changes are needed. But their criticism of the standards was mainly directed towards the insufficient reflection of the technological advancement in the connected software development processes. The article does not cover specific characteristics of autonomous vehicles and the ecosystems in which they operate. Other criticisms of certification procedures were detected, too. In \citep{burzio2018}, which asks for modifications from the standpoint of cyber-security. But the most often addressed certification problems are those related to the safety of AVs, as presented in \citep{zhao2022} or \citep{cummings2019}. The last mentioned paper specifically criticizes possibly lower vehicle safety unless software upgrades are taken into account in the certification process.

Naturally, the identified shortcomings and criticism of the currently used certification procedures sparked the development of suggestions for improvements. In \citep{unece2022}, GRVA\footnote{Working Party preparing draft regulations, guidance documents and interpretation documents for adoption by the parent body} group working under United Nations Economic Commission for Europe (UNECE) in collaboration with experts of the International Organization of Motor Vehicle Manufacturers (OICA) presented a new way of validating autonomous vehicles for the purpose of certification based on a multi-pillar approach consisting of a scenario catalog. \citet{dynamic2022} proposed dynamic certification built on modelling and testing, which are constantly intertwined during the system's life cycle. Besides that, a verifiably-correct dynamic self-certification framework for autonomous systems is discussed in \citep{fisher2018}, while \citet{hussein2021} introduced the concept of a certification framework for autonomous driving systems based on the Turing test. Digital certificates are proposed to be used in combination with trust and reputation policies for ensuring safety and detection of hijacking vehicles in \citep{garcia2019}. All of the attempts partially cover the shortcomings of static certification, however, the debate regarding certification is grounded solely in the context of safety and security and lacks to consider other aspects, such as ethics and trust of AVs. 

% trust and ethics considerations - standards yes, certification not yet
Even though trust and ethics are in general frequently debated issues in the context of self-driving cars,
% despite pointing out their importance \citep{henschke2020,trust-cioroaica2020}, 
the literature presents these concepts more in connection with privacy preservation \citep{lai2021}, or general calls for adjusting the software development standards and best practices \citep{kwan2021,myklebust2020} for building software with social responsibility. Considering trust and ethics directly within the certification process itself does not seem to be covered yet. 
To this end, if we are to trust the systems that make important decisions for us not only in the area of safety, but also in ethics, we must ensure that these aspects are also considered at the stage of vehicle certification providing legal guarantees.

\section{\uppercase{Specification of the certification context}}
\label{sec:specs}
Responsible system certification can be hardly achieved when performed both in isolation from the environment in which the system operates, and without taking into account the characteristics of the system under consideration.
AVs are considered cyber-physical systems (CPSs), i.e. \textit{”smart networked systems with embedded sensors, processors and actuators that are designed to sense and interact with the physical world”} \citep{cps-def}. Therefore, the characteristics of CPSs can aid with the identification of the characteristics of AVs needed for thorough certification. 

The aim of this section is to cover the key characteristics of autonomous CPSs as well as the whole surrounding ecosystems, and specify the certification context based on which we then define the requirements for certification standards for future autonomous systems.

\subsection{Autonomous cyber-physical systems}
The characteristics of autonomous systems need to be considered when determining requirements for certification frameworks for driving systems.
\citet{weyns2021} defines key principles for future CPS engineering principles, from which the characteristics of future autonomous systems can be derived. The following principles are listed: 
(1)~\textit{crossing boundaries} related to close contact between social-, physical-, and cyber-spaces;
(2)~\textit{leveraging the humans} and their integration in the design and operation processes instead of treating them only as users of a system;
% (3) fluid modelling
(3)~\textit{on-the-fly coalitions} as a way of addressing complex problems through forming multi-agent systems;
(4)~\textit{dynamically assured resilience} to withstand uncertainty, context changes or any other disruptions and continue in service provision;
(5)~\textit{learn novel tasks}, that is, utilizing knowledge from the past effectively to deal with novel situations.

\begin{figure*}[t]
\centering
\includegraphics[width=330px]{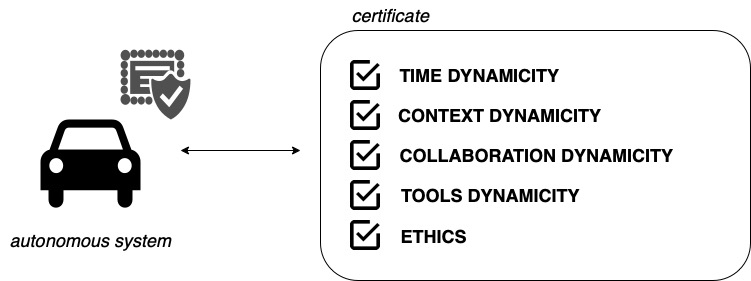}
\caption{Key certification standards' aspects on future autonomous systems}
\label{fig:reqs}
\end{figure*}

\subsection{Dynamic software ecosystems}
Besides that, the ecosystem in which the autonomous system operates must also be considered. The ecosystem identifies its surroundings and entities communicating with the system, specifies the relationships between them as well as determines the system context, that is, the system's main meaning and ways of use.

\citet{deco2021} defined ecosystems formed by software autonomous systems as ecosystems supporting dynamic, smart, and autonomous features which are required by modern software systems. In particular, the following features need to be taken into consideration when designing certification suitable for autonomous systems: 
(1) \textit{automation} understood as automated self-adaptation of the ecosystem on context changes during runtime;
(2) \textit{autonomy} meaning there is no longer any human monitoring of the systems inside the ecosystem; 
(3) \textit{dynamic goal evaluation} as intelligent adaptation to dynamic needs  with the intention to achieve the proposed and explicitly defined goals; 
% (4) \textit{goals classification} as support for self-regulating mechanisms; and 
(4) \textit{automated trust management} as a crucial concept driving decision-making\footnote{the overarching ecosystem goal derives from the achievement of tactical goals on lower levels and collaboration, which is needed to fulfill the main goal, relies on trust guarantees in such dynamic and hard-to-predict environments}; and 
(5) \textit{architecture implications} as the ecosystem's dynamic nature, driven by the constant (dis-)connecting nodes, impacts the architecture of the ecosystem.

\subsection{Ethical aspects}
In case of human-operated vehicles, the drivers themselves are responsible for applying the basic rules of safety and morality when driving. However, integrating autonomous systems into ecosystems shared with humans shifts the ethical aspects of driving on the system. And when humans are not in charge anymore, the system must bear a certain moral obligation instead. 

Examples of such circumstances include the well-known trolley problem \citep{trolley}, in which a collision is irreversible and any action that is made results in a tragedy. Other examples where ethical aspects are applied concern an exhibition of altruistic behaviour (e.g., informing other road members about a danger they cannot yet see, such as a person crossing the street on the red light), or a demonstration of solidarity with surrounding entities (e.g., transparency in terms of notifying other drivers of one's change in speed or direction before doing so, in order to maintain smooth traffic and avoid dangerous situations).

Thus besides the primary requirements on autonomous systems, which is assuring safety and security \citep{safety-and-security-first}, ethical considerations might govern the gray space where it becomes clear that some level of harm might happen anyway, or an action can be done to help the overall ecosystem, although such action is not required by law.

% In case of human-operated vehicles, the drivers themselves are responsible for applying the basic rules of safety and morality when driving, such as moving out of the way for a rushing emergency vehicle, not parking in parking slots reserved for people with disabilities if not being eligible for it, or stopping for pedestrians crossing the street in a crosswalk. However, integrating autonomous systems into ecosystems shared with humans shifts the ethical aspects of driving on the system. And when humans are not in charge anymore, the system must bear a certain moral obligation instead. 

% % Thus besides the primary requirements on autonomous systems, which is safety and security, ethical considerations might govern the gray space where we either know that some level of harm might happen anyway, or we might do an action to help the overall ecosystem although we do not have to.

% The aim of this paper is not to go against the primary requirements placed on autonomous systems, which is to assure safety and security in every situation. However, we are convinced that it is essential that ethical considerations become a standard principle of autonomous vehicles, not that it is considered only as a nice-to-have feature.
% As stated above, an autonomous CPS system does not operate independently but shares the ecosystem with other entities, such as technological devices but also people, and impacts them both substantially. Therefore, ethical principles must be considered.

The idea of implementation of ethical principles in autonomous driving elaborates even further on one of the requirements for a CPS, in particular \textit{leveraging the humans}. When ethical standards are applied, people are no longer seen as just users but as equal members of the ecosystem with individual needs and dignity.

The research generally agrees on the fact that applying ethical norms is crucial for trust formation in autonomous systems \citep{towards-kwan-2021,ethics-wang-2022,virtue-gerdes-2020}. In human society, ethical diversity is observed, which necessitates the possibility to adjust the moral setting of a system according to one's individual preferences within a personal ethical framework. However, a personal ethical framework with no regulation is insufficient as it would lead to numerous ethical dilemmas. A preferred ethical model should therefore reflect public rational ethical inclinations while at the same time provide users with limited freedom to adjust the particular ethical setting according to their personal preferences \citep{ethics-wang-2022}. 
The contribution to the discussion regarding the form and way of the implementation of specific ethical principles in autonomous systems is out of the scope of this paper, though. Instead, as this paper deals with the readiness of current certification standards for autonomous vehicles, we focus on the examination of whether individual standards take into account ethical issues regarding the driving task.

\section{\uppercase{Methodology}}
To identify the gaps within the certification procedures of autonomous systems, the requirements on these systems need to be defined and a set of standards for evaluation needs to be established. We derive the requirements based on the characteristics of autonomous systems as well as based on the characteristics in which such systems operate. We draw the knowledge from literature devoted to the study of autonomous systems in dynamic software ecosystems. As for the standards selection, we have selected standards published or re-confirmed within the last seven years (2016-2022) in various fields of the automotive domain from international organizations for standardization and evaluated their readiness for fully autonomous driving. The steps are more thoroughly described below.

\begin{figure*}[t]
\centering
\includegraphics[width=420px]{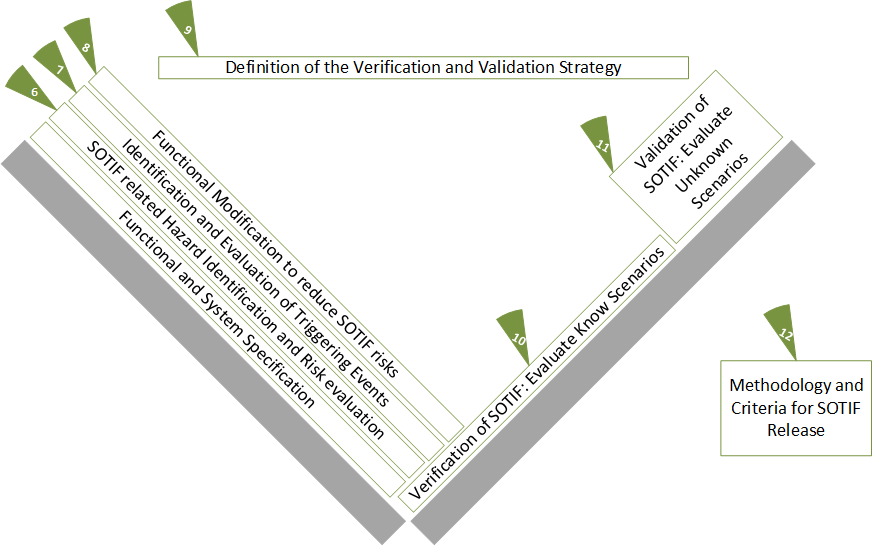}
\caption{Elements of safety of the intended functionality defined by the standard ISO 21448 \citep{iso-figure-source}}
\label{fig:iso}
\centering
\end{figure*}

\subsection{Requirements selection}
The characteristics of the dynamic software ecosystems, in which autonomous systems operate, as well as the systems themselves, presented in Section \ref{sec:specs}, lead to the need for the shift from staticity towards a more dynamic approach.
By staticity, we mean the certification standards that shall no longer be static in terms of:

\bigskip
\begin{enumerate}
    \item \textit{time}, that a vehicle is granted a certificate based on its status at a single time point, typically at the moment of production,
    \item \textit{context}, that the certificate is granted based on pre-defined and finite set of tests,
    \item \textit{collaboration}, that the standard neither supports dynamic creation of coalitions nor considers communication with other entities in the environment,
    \item static \textit{tools} used within the certification process, such as documentation examination, model report assessment or visual inspection.
\end{enumerate}

As we believe future certification standards should be revised to reflect the ecosystems' dynamicity, we suggest the shift the aforementioned static characteristics into their dynamic version, towards which we evaluate the selected standards to check their readiness for autonomous systems. We refer specifically to the dynamicity in the following aspects:

\begin{enumerate}
    \item \textit{time}, that the standard is able to deal with dynamic system changes, e.g. caused by a software update,
    \item \textit{context}, that the standard suggests using tools to thoroughly verify the systems' functionality in dynamic context, such as unforseen and upredictable situations,
    \item \textit{collaboration}, that the standard supports dynamic creation of multi-agent systems to enable optimization and solving of complex problems more efficiently, e.g. through on-the-fly coalitions,
    \item \textit{tools}, to check whether the tools used to verify the systems' compliance with the standard are dynamic.
\end{enumerate}

For the analysis of the selected standards, in addition to the dynamic characteristics, we also consider

\begin{enumerate}
  \setcounter{enumi}{4}
  \item \textit{ethics}, meaning whether ethical issues are explicitly discussed or taken into account within the cases addressed in the particular standard under evaluation.
\end{enumerate}

The selected key certification standards aspects on future autonomous systems are visualized in Figure \ref{fig:reqs}.

\subsection{Standards selection}
\label{sec:standard-selection}
The documents for evaluation were chosen in the following manner. First, we seached for standards issued or re-confirmed in the last seven years (2016-2022) by international organizations for standardization (ASAM, ISO, ITU, SAE, UNECE). We focused on standards published for light-duty vehicles (passenger cars) in the domain of on-road automated driving, regarding software requirements specification, software updates, testing, validation and verification, or guidance for automated driving in general. 

From this search, we identified 27 relevant documents. However, not all of the selected items could be analyzed. In addition to documents publicly unavailable at the time of writing of this paper (e.g. ISO/AWI TS 5083 being under development), we also omitted documents defining only the taxonomy and terminology, exchange format or language specifications (ASAM standards), and documents not directly related to automated driving systems (e.g. audit and other control activities guidelines, such as in ISO/PAS 5112). The final list of standards selected for evaluation is in Table \ref{table:description}.

\section{Standards evaluation}
% static - time (single point of certification), context (predefined tests), goals (safety, security, emissions), tools (documentation examination, specification checks, model report assessment, visual inspection)
After the selection of relevant documents, we performed an evaluation of the standards' suitability for fully autonomous driving against the requirements defined in Section \ref{sec:standard-selection}. Most of the selected standards are concerned with safety, which is understood in terms of a) functional and system specification, b) identification of triggering events, c) reduction of safety risk, and d) validation and verification of the system's functionality (see Figure \ref{fig:iso}).

The evaluation results are presented in Table \ref{table:evaluation}, the standards are sorted according to their identifier in alphabetical order. The evaluation results in each of the reviewed aspects are discussed in a more detailed manner in the following paragraphs, with providing the evaluation summary at the end of this section.

\subsection{Time dynamicity}
Only seven of the reviewed documents show readiness for the dynamicity in the time context. That is, these standards include measures for the case when there are additional updates of an already certified system, and do not require re-execution of the whole certification process conducted by an authority. Typically, these standards provide guidelines for creating the necessary documentation of changes, performing thorough testing, and arranging on-road monitoring for verifying that the update has not disrupted the intended functionality of the vehicle (UL 4600). 

The rest of the standards do not provide any information regarding software updates (e.g. ISO 15037, ISO 22735). This makes it unclear how to proceed in case changes in the software are needed, such as ensuring recognition of a new type of traffic sign, and whether the awarded certificate becomes invalid after an update in the already certified software.

\subsection{Context dynamicity}
Unlike all other categories, dynamicity within the context category seem to be relatively well addressed. In terms of system verification, most of the standards require performing simulation tests with unforseen and upredictable situations to thoroughly verify the systems' safety or security in the development phase of an autonomous vehicle system. Some standards, such as ISO 21448, provide detailed guidance on verifying the system's performance in unknown scenarios. Public road testing is also widely used in the reviewed documents as a testing method in the final stages of the development process. However, techniques and requirements for public road-testing rules differ over countries, and a global regulatory framework for public road AV testing is yet to be developed~\citep{bakar2022}. 

Standards not supporting context dynamicity, which are missing a check mark in the table, usually rely on the verification of systems' functionality based only on predefined set of tests (ISO 11270, SAE J3018), or the information about testing tools is missing due to the standard's focus area (ISO/TS 5255).

\begin{table*}[htp]
\footnotesize
\centering
\begin{tabular*}{\textwidth}{c l l l}
\hline
\textbf{\hspace{5pt}No.\hspace{5pt}} & \textbf{Standard ID \hspace{50pt}} & \textbf{Name \hspace{180pt}} & \textbf{Field}    \\ 
\hline
\hline

1            & ISO 11270         & \makecell[l]{Intelligent transport systems - \\Lane keeping assistance systems (LKAS) — \\Performance requirements and test procedures} & Safety\&Assurance \\ \hline
2            &    ISO 15037 series    &     \makecell[l]{Road vehicles — Vehicle dynamics test methods}                                                                                                                  &   Safety\&Assurance                \\ \hline
3             &       ISO 21448    &      \makecell[l]{Road vehicles — Safety of the intended functionality}                                                                                                                 &    Safety\&Assurance               \\ \hline
 4            &      ISO 22735            &    \makecell[l]{Road vehicles — Test method to evaluate \\the performance of lane-keeping assistance systems}                                                                                                                   &          Safety\&Assurance         \\ \hline
  5           &      ISO 22737             &      \makecell[l]{Intelligent transport systems — \\Low-speed automated driving (LSAD) systems for \\predefined routes — Performance requirements, \\system requirements and performance test procedures}                                                                                                                 &       Safety\&Assurance            \\ \hline
 6            &      ISO/DIS 26262 series             &  \makecell[l]{Road vehicles — Functional safety}                                                                                                                     &     Safety\&Assurance              \\ \hline
  7           &      ISO/SAE 21434             &         \makecell[l]{Road vehicles — Cybersecurity engineering}                                                                                                              &                  Cybersecurity \\ \hline
  8           &    ISO/TR 21959 series               &        \makecell[l]{Road vehicles — Human performance and state \\in the context of automated driving}  &   Human factor     \\ \hline
  9           &  ISO/TR 4804      &     \makecell[l]{Road vehicles — Safety and cybersecurity for \\automated driving systems — \\Design, verification and validation}    &    \makecell[l]{Safety\&Assurance, \\Cybersecurity}   \\ \hline
  10           &   ISO/TS 5255 series     &     \makecell[l]{Intelligent transport systems — \\Low-speed automated driving system (LSADS) service}    &   Data    \\ \hline
  11           &   SAE J2945     &     \makecell[l]{On-Board System Requirements for \\V2V Safety Communications}    &  Safety\&Assurance     \\ \hline
  12           &   SAE J3048     &     \makecell[l]{Driver-Vehicle Interface Considerations \\for Lane Keeping Assistance Systems}    &   Safety\&Assurance    \\ \hline
  13           &   SAE J3061     &     \makecell[l]{Cybersecurity Guidebook for \\Cyber-Physical Vehicle Systems}    &   Cybersecurity    \\ \hline
  14          &   UL 4600     &     \makecell[l]{Standard for Safety for the \\Evaluation of Autonomous Products}    &    Safety\&Assurance   \\ \hline
\end{tabular*}
\caption{List of standards selected for evaluation.}
\label{table:description}

\bigskip
\bigskip
\bigskip

\centering

\begin{tabular}{c cccc c }
\hline
% \multirow{2}{*}{\hspace{5pt}\textbf{No.}\hspace{5pt}} & \multicolumn{4}{c }{\textbf{Dynamicity in}}                                                                & \multirow{2}{*}{\hspace{12pt}\textbf{Ethics}\hspace{12pt}}  \\ 
% \cline{2-5}
%                           & \multicolumn{1}{c }{\hspace{5pt}\textbf{Time}\hspace{15pt}} & \multicolumn{1}{c }{\hspace{15pt}\textbf{Context}\hspace{15pt}} & \multicolumn{1}{c }{\textbf{Collaboration}} & \multicolumn{1}{c }{\hspace{5pt}\textbf{Tools}\hspace{15pt}} &                          \\ 

\hline
\makecell[c]{ \textbf{\hspace{5pt}No.\hspace{5pt}} } & \makecell[c]{ \textbf{Time dynamicity} } &  \hspace{5pt}  \makecell[c]{ \textbf{Context dynamicity} } & \hspace{5pt} \makecell[c]{ \textbf{Collaboration dynamicity} } & \hspace{5pt} \makecell[c]{ \textbf{Tools dynamicity} } &  \hspace{5pt} \makecell[c]{ \textbf{Ethics} }
\\ 

\hline
\hline

1 &   \multicolumn{1}{c }{-} & \multicolumn{1}{c }{-}   & \multicolumn{1}{c }{-}  & \checkmark &   -  \\ \hline
2 &   \multicolumn{1}{c }{-} & \multicolumn{1}{c }{ \checkmark }   & \multicolumn{1}{c }{-}  & - & - \\ \hline
3 &   \multicolumn{1}{c }{ \checkmark } & \multicolumn{1}{c }{ \checkmark }   & \multicolumn{1}{c }{ \checkmark }  & \checkmark &  \checkmark \\ \hline
4 &   \multicolumn{1}{c }{-} & \multicolumn{1}{c }{ \checkmark }   & \multicolumn{1}{c }{-}  & - & -  \\ \hline
5 &  \multicolumn{1}{c }{-} & \multicolumn{1}{c }{ \checkmark }   & \multicolumn{1}{c }{-}  & - & -  \\ \hline
6 &   \multicolumn{1}{c }{ \checkmark } & \multicolumn{1}{c }{ \checkmark }   & \multicolumn{1}{c }{ \checkmark }  & \checkmark &  - \\ \hline
7 &   \multicolumn{1}{c }{ \checkmark } & \multicolumn{1}{c }{ \checkmark }   & \multicolumn{1}{c }{ \checkmark }  & \checkmark &  - \\ \hline
8 &  \multicolumn{1}{c }{-} & \multicolumn{1}{c }{ \checkmark }   & \multicolumn{1}{c }{-}  & - & \checkmark  \\ \hline
9 &   \multicolumn{1}{c }{ \checkmark } & \multicolumn{1}{c }{ \checkmark }   & \makecell[c]{ unclear collaboration \\form and purpose }  & \checkmark &  \checkmark \\ \hline
10 &   \multicolumn{1}{c }{-} & \multicolumn{1}{c }{-}   & \makecell[c]{ unclear collaboration \\form and purpose }  & - & -  \\ \hline
% ITU FGAI4AD series &  Safety\&Assurance & \multicolumn{1}{c }{} & \multicolumn{1}{c }{ covered }   & \multicolumn{1}{c }{}  & covered &  covered  \\ \hline
11  &   \multicolumn{1}{c }{\checkmark} & \multicolumn{1}{c }{-}   & \makecell[c]{ unclear collaboration \\form and purpose }  & \checkmark & \checkmark \\ \hline
12  &   \multicolumn{1}{c }{-} & \multicolumn{1}{c }{-}   & \makecell[c]{ unclear collaboration \\form and purpose }  & - & - \\ \hline
13 &   \multicolumn{1}{c }{ \checkmark } & \multicolumn{1}{c }{ \checkmark }   & \multicolumn{1}{c }{-}  & \checkmark & -  \\ \hline
14 &   \multicolumn{1}{c }{ \checkmark } & \multicolumn{1}{c }{ \checkmark }   & \makecell[c]{ unclear collaboration \\form and purpose }  & \checkmark &  \checkmark \\ \hline
\end{tabular}
\caption{Evaluation of certification standards. The numbers of standards match with the list of standards defined in Table \ref{table:description}.}
\label{table:evaluation}
\end{table*}

\subsection{Collaboration dynamicity}
As for the collaboration dynamicity, standards ISO/DIS 26262 and ISO 21448 provide a comprehensive specification and design of communication between vehicle and other entities within the surrounding ecosysystem. Besides that, ISO/SAE 21434 incorporates distributed cybersecurity activities, assigning cybersecurity responsibilities between multiple parties. Because these three standards explicitly address the possibility of collaboration of multiple agents within the ecosystem, they were marked as fully supportive towards dynamic collaboration. 

However, even though the dynamic creation of coalitions is covered by these documents, we still see an opportunity for improvement. We notice that none of the reviewed standards considers any form of trust management when collaborating with other entities. Since the entities might have malicious intentions, naively trusting any entity willing to collaborate poses a serious security risk to the operation of the entire ecosystem. 

In the UL 4600, ISO/TR 4804, ISO/TS 5255, SAE J2945 and SAE J3048 standards, terms like \textit{"dependencies between items"}, \textit{"vehicle to vehicle communication"}, or \textit{"arrays of systems implementing other vehicle level functions"} have been found. Thus, the possibility of vehicle communication with other entities within their surrounding environment is at least partially addressed. However, it is unclear (1)~how and in which directions the communication is carried, and (2)~whether joint strategy creation is possible. Because of that, these standards were marked as providing only a partial support. All other standards do not seem to directly support collaboration dynamicity at all, or just to a negligible extent.

\subsection{Tools dynamicity}
More than a half of the evaluated documents enable using of dynamic tools for checking the system's compliance with a given standard. Such a check of compliance typically includes on-site testing or execution of further system's verification and validation measures. Even in the cases where dynamic tools are present, a large part of the compliance checks is still performed by static tools, such as manual inspection of documentation, processes and procedures. Standards which are not identified as supportive towards the use of dynamic tools in the assessment table employ only static tools for compliance checks.

\subsection{Ethics}
The series of ISO/TR 21959 standards is devoted to ethics in a large extent. Except for the ethical considerations, the standards also present the best practices in the field of societal trust formation. In particular, they establish guidelines for a better acceptance of autonomous vehicles based on the analysis of various human factors having an impact on perception and trust formation towards autonomous systems within the automotive domain.

But ethical considerations do not appear to be generally taken into account by the assessed standards, though. Four standards in the Safety\& Assurance area refer to the need of absence of unreasonable risk, which is defined as a risk that is \textit{"unacceptable in a certain context according to valid societal moral concepts"} \citep{iso21448}. Other moral aspects do not seem to be addressed, and we find the absence of consideration of moral aspects particularly problematic. We are convinced that autonomous systems such as autonomous vehicles, which have direct responsibility for human lives, bear moral obligation.

The ethics of AI systems in general is a topic that is currently frequently debated, and the authorities are working on complex market regulation strategies. The activities of the European Commission may serve as an illustration. Starting with the publication \textit{"Policy and Investment Recommendations for Trustworthy Artificial Intelligence"} \citep{trustworthy-ai-2019}, the European Commission laid the foundations of AI regulation in the European Union, covering the fundamental questions on the border of law and ethics. Since then, multiple documents discussing ethics have been published, but many concerns and questions are still unanswered. 

Since ethics is a complex issue, it is possible that the authors of standards in the field of autonomous driving are awaiting the central regulatory bodies' further recommendations and will adjust the documents once the direction in which the regulation will evolve is clear. In any case, autonomous systems cannot be developed without ensuring human and societal needs are taken into consideration during the systems' operation. We are convinced that even on a technical level, it is necessary to establish mechanisms which can control the observance of ethical principles, and which can be followed up after further guidelines are published by the central authorities.

\subsection{Evaluation summary}
To sum up, the evaluation results are diverse. While some of the standards show clear insufficient preparedness to reflect the dynamicity of future autonomous systems and ecosystems in which they operate (e.g. ISO 22735, ISO 22737, or SAE J3018), other standards show signs of readiness for autonomous driving after meeting all (ISO 21448) or a vast majority (ISO/TR 4804, UL 4600) of the specified requirements. Yet, there is room for improvement even in this case.

Namely, in all the analyzed standards, we identified the absence of considering the creation of dynamic coalitions as well as the absence of trust management in standards with support towards coalition dynamicity as one of the most urgent deficiencies from the evaluated aspects. Besides that, we undoubtedly see shortcomings in the field of ethics, too. We observe that most of the standards are narrowly focused, usually towards safety, while neglecting ethical aspects, which are often directly related.

% Future adjustments of the certification processes are unavoidable. Vehicles are expected to become even more complex, and overtake all critical driving functions. 

% Current certification methods seem not to be designed to promptly deal with the changes, though. Inspired by the analysis performed by \citep{fisher2018,dynamic2022}, we have identified the staticity of the currently used certification approaches as the most critical issue for the future. By staticity, we mean the current approaches are static in terms of (1) \textit{time}, as a vehicle is granted a certificate based on its status at a single time point, typically at the moment of production; (2) \textit{context}, as the certificate is granted based on pre-defined and finite set of tests (3) \textit{goals}, as the certificates are issued towards a single and unchangeable objective (e.g. safety, security, emissions); and (4) \textit{tools} used within the certification process, such as documentation examination, model report assessment or visual inspection. 

\section{\uppercase{Suggested Improvements}}
\label{sec:solution}
Given the presented evaluation and a relatively long process of preparation and approval of standards, which typically takes three years from the initial proposal to its final publication according to ISO\footnote{\url{https://www.iso.org/developing-standards.html}}, it is not surprising that standards do not fully reflect the advancement in dynamic autonomous systems. Several shortcomings in the field of ethics were identified, too. Yet, there are mechanisms that could help the standards better support the dynamicity of the software ecosystems and the future autonomous cyber-physical systems, while reflecting the moral setting of the society as well. 

In this section, we outline five suggested directions that could lead to a better fit of the standards with the requirements of future dynamic autonomous systems and ecosystems, including an advanced assurance of ethical awareness. 

\subsection{Real-time validation of a certificate and its properties}
Traditional certificates, whose purpose is to provide certain guarantees about the quality of the certified system, are granted at a specific point in time, usually at the time of production of the certified system. However, in a dynamic and ever-changing environment, such a certificate may not keep its validity over time; instead, its validity may change depending on the context or deteriorate over time.

The certification schemes for dynamic autonomous systems shall account for some form of identification or quantification of the certificate deterioration, in order to address the issue of time dynamicity. In other words, in a dynamic environment, it is necessary to constantly re-check the certificate's validity and decide to what extent the certificate and its guarantees can be trusted.

For better illustration of the need for the implementation of real-time certificate validation, consider a scenario where an AV has a valid certificate (\textit{valid} in the traditional meaning of the certification concept) but its behavior is suspicious, raising the possibility that the vehicle has malicious intentions. Other vehicles will be able to react to such a situation more effectively if there are mechanisms for evaluating suspicious manners that deviate from the expected, typical behavior. In this case, the real-time validity check would show that the vehicle was not certified to handle this specific situation, or that the certificate is outdated because vehicle's software has not been updated on a version fixing critical bugs that allow attackers to hack the vehicle. 

\subsection{Certificate combined with vehicle's reputation}
The examination of certification standards revealed certain flaws, which might lead to a false sense of security. As discussed above, there may be situations in which the vehicle is certified, yet its safe operation is impaired (e.g. after a faulty software update that is not detected by static tools used to check the system's compliance with a standard). A~possible solution to this issue could be linking the certificate to the vehicle's reputation.

In dynamic ecosystems, deciding who to trust becomes a challenging task. As studied in the disciplines of trust, reputation is used as one of the tools for trust-building. Reputation can be defined as \textit{the overall quality of an entity derived from the judgements by other entities in the underlying network, which is globally visible to all members of the network} \citep{trust-and-reputation}, and when such information is propagated through the network of connected entities, it can have a substantial effect on decision-making. Besides that, by providing information allowing distinction between trustworthy and untrustworthy nodes, reputation can also help in dealing with observable misbehavior \citep{rep-def} and minimizing damage in case of an insider attack \citep{reputation1}. 

This concept may be particularly useful for addressing the time dynamicity problem within certification. Same as people gain reputation by having their actions evaluated by their peers, an entity's reputation in a smart ecosystem is dependent on how it behaves and interacts with others \citep{buhnova2022tutorial}. Moreover, a collection of experience during runtime feeding updates of the score of trustworthiness could also be used to promote or demote the certificate's validity. 

Tying the certificate to the vehicle’s reputation could help other entities respond to changes in a dynamic environment in an even more flexible manner. Demonstrated on a hypothetical scenario, a vehicle would be less trusted if its reputation reported by other vehicles declined despite having a valid certificate, caused by other vehicles reporting its sudden suspicious behavior (indicating a software bug or an attack). In case of a serious reputation drop, the vehicle's certificate could be temporarily or completely revoked to prevent further damage.

\subsection{Extension of the certificate's status to a scale}
Another alternative, which is partly related to the linkage of the certificate to the concept of reputation discussed above, is to redefine the concept of certification in terms of the range of potential values it can acquire. The traditional view on a certificate nowadays treats it as a binary value (\textit{valid/invalid}, or \textit{granted/not granted} certificate). Such a perception might be unnecessarily restricting. 

Instead, we suggest rethinking the concept and considering it rather a scale to better represent the current certification status of the system installed in the vehicle or other autonomous system. Indeed, there are multiple ways to interpret this newly proposed concept. To mention some of the possible meanings, the scale might represent e.g. the number of software updates installed in the system, or the amount of time elapsed since the last official verification of the system's compliance with a particular standard. The exact interpretation of the certification scale is up for further research and discussion.

\subsection{Considering certificate's context-dependant validity}
The implementation of trust management appears to be necessary even in addressing the context and coalition dynamism issue. During runtime, evaluating the acceptable level of trust, or eventually the vulnerability risks, is strongly context-dependent.

Consider two AVs initializing mutual communication in two scenarios. In scenario A, the AVs' intention is to exchange a batch of weather-related data and then stop any further interaction. In scenario B, the entities eshablish a connection with the purpose of creating a vehicle platoon in order to reduce fuel consumption due to lower air resistance. However, close collaboration needed for vehicle platoon formation in Scenario B raises more serious trust concerns about the safety of riding in such close proximity compared to the situation of exchanging data in Scenario A. And even if a vehicle may be certified in correctly handling interaction with other vehicles in some contexts, its behavior may not be verified or guaranteed in the other contexts. Therefore, before engaging in any interaction, it is the system's responsibility to verify whether the awarded certificate, as well as the other party, can be trusted in the given context.

\subsection{Certificates combined with ethical concerns}
Evaluating the safety of products or their environmental footprint before they are allowed to enter the market is nowadays a common practice. But the review of current standards in this paper has shown that ethical aspects are still not frequently considered. Technology development has to be ethically supervised, though. Otherwise, intelligent systems developed with the intention to help can easily turn to harm or disadvantage certain groups of people.

Our idea to address the certification gap regarding the ethics of autonomous systems is to assess the ethics of an autonomous system in the same way as safety or environmental aspects. In particular, we suggest combining certification with Ethical Digital Identities (EDI), a concept introduced by \citet{edi2022}. Derived from the concept of Digital Identities \citep{digital-identities-2005}, EDI serve as the basis for safeguarding the evolution of intelligent safety-critical systems in terms of ethics.

\section{\uppercase{Conclusion}}
\label{sec:conclusion}
In this paper, we evaluated the readiness of current certification standards for future autonomous driving systems. We analyzed the characteristics of both autonomous systems and the dynamic software ecosystems in which they operate, from which we derived a set of requirements, namely \textit{Time Dynamicity, Context Dynamicity, Collaboration Dynamicity, Tools Dynamicity}, and \textit{Ethics}, which we then used to assess the standards.

The results demonstrate that the present standards are not entirely ready for the expansion of autonomous driving systems, and also assisted us in identifying their primary shortcomings. One of the most serious deficiencies is referring to the \textit{Collaboration Dynamicity} aspect. We criticize mainly the lack of support for the creation of dynamic coalitions among standards, as well as the complete absence of any kind of trust management strategies for establishing communication with other entities. Another shortcoming concerns neglecting ethical aspects in the standards' focus.

In order to address the identified gaps, we outlined a concept for the improvement of certification standards. We present five ideas for rethinking the certification that we believe will help move the discussion towards a complete solution of the identified problems, so that standardization for autonomous systems (not only those in the automotive domain) will better fit the requirements of future dynamic systems and ecosystems with ethical awareness, which can be trusted. The presented ideas are subject for further research and will be elaborated on in our future work.

\section*{\uppercase{Acknowledgements}}

This research was supported by ERDF "CyberSecurity, CyberCrime and Critical Information Infrastructures Center of Excellence" (No. CZ.02.1.01/0.0/0.0/16\_019/0000822).

%   The authors would also like to thank the anonymous referees for
%   their valuable comments and helpful suggestions. The work is
%   supported by the \grantsponsor{GS501100001809}{National Natural
%     Science Foundation of
%     China}{http://dx.doi.org/10.13039/501100001809} under Grant
%   No.:~\grantnum{GS501100001809}{61273304}
%   and~\grantnum[http://www.nnsf.cn/youngscientsts]{GS501100001809}{Young
%     Scientsts' Support Program}.

% If any, should be placed before the references section
% without numbering. To do so please use the following command:
% \textit{$\backslash$section*\{ACKNOWLEDGEMENTS\}}

\bibliographystyle{apalike}
{\small
\bibliography{example}}

% \section*{\uppercase{Appendix}}

% If any, the appendix should appear directly after the
% references without numbering, and not on a new page. To do so please use the following command:
% \textit{$\backslash$section*\{APPENDIX\}}

\end{document}